\documentclass[manuscript,screen,nonacm]{acmart}
\acmConference[]{} 

\newcommand{\workshopname}{GenAICHI: CHI 2024 Workshop on Generative AI and HCI}
\newcommand{\licensedetails}{Licensed under a Creative Commons Attribution 4.0 International License (CC BY 4.0). Copyright remains with the author(s).}
\newcommand\extrafootertext[1]{
    \bgroup
    \renewcommand\thefootnote{\fnsymbol{footnote}}%
    \renewcommand\thempfootnote{\fnsymbol{mpfootnote}}%
    \footnotetext[0]{#1}%
    \egroup
}

\AtBeginDocument{ 
    \fancypagestyle{firstpagestyle}{
        \fancyhf{}
        \fancyfoot[L]{\sffamily\footnotesize \workshopname}%
        \fancyfoot[C]{\sffamily\footnotesize \thepage}
    }
    \fancyhf{}
    \fancyhead[L]{\sffamily\footnotesize\shorttitle}
    \fancyhead[R]{\sffamily\footnotesize\shortauthors}
    \fancyfoot[L]{\sffamily\footnotesize\workshopname}%
    \fancyfoot[C]{\sffamily\footnotesize\thepage}
    \extrafootertext{\licensedetails}
}
\usepackage{lipsum} 

\usepackage{enumitem}
\setlist{nosep}
\usepackage{titlesec}
\begin{document}

\title{Equivalence: An analysis of artists’ roles with Image Generative AI from Conceptual Art perspective through an interactive installation design practice}

\author{Yixuan Li}
\affiliation{%
  \institution{Georgia Institute of Technology}
  \city{Atlanta}
  \country{USA}
}
\author{Dan C. Baciu}
\affiliation{%
  \institution{Delft University of Technology}
  \city{Delft}
  \country{Netherlands}
}
\author{Marcos Novak}
\affiliation{%
  \institution{University of California, Santa Barbara}
  \city{Santa Barbara}
  \country{USA}
}
\author{George Legrady}
\affiliation{%
  \institution{University of California, Santa Barbara}
  \city{Santa Barbara}
  \country{USA}
}

\renewcommand{\shortauthors}{Yixuan Li et al.}

\begin{abstract}
Over the past year, the emergence of advanced text-to-image Generative AI models has significantly impacted the art world, challenging traditional notions of creativity and the role of artists.
This study explores how artists interact with these technologies, using a 5P model (Purpose, People, Process, Product, and Press) based on Rhodes' creativity framework to compare the artistic processes behind Conceptual Art and Image Generative AI. To exemplify this framework, a practical case study titled "Equivalence", a multi-screen interactive installation that converts users' speech input into continuously evolving paintings developed based on Stable Diffusion and NLP algorithms, was developed. Through comprehensive analysis and the case study, this work aims to broaden our understanding of artists' roles and foster a deeper appreciation for the creative aspects inherent in artwork created with Image Generative AI.
\end{abstract}



\begin{teaserfigure}
  \includegraphics[width=\textwidth]{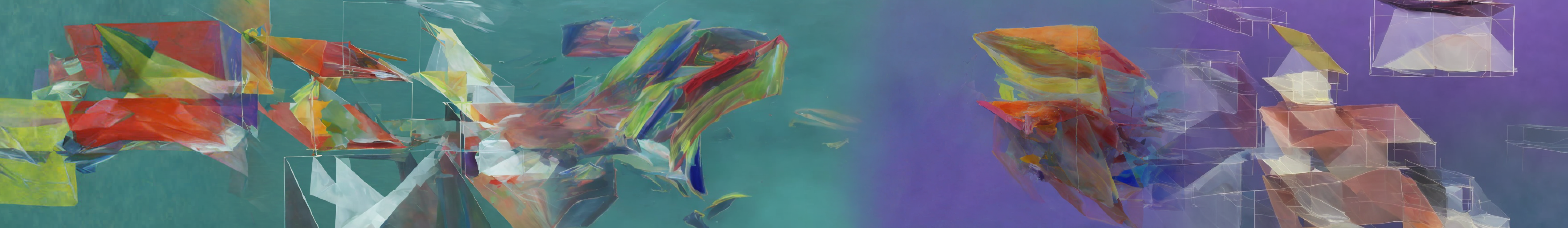}
  \caption{Screenshot from Equivalence: An audio interactive installation visualization speech into structures in space}
  \Description{Screenshot from Equivalence: An audio interactive installation visualization speech into structures in space}
  \label{fig:teaser}
\end{teaserfigure}

\received{20 February 2007}
\received[revised]{12 March 2009}
\received[accepted]{5 June 2009}

\maketitle

\section{Introduction}
Generative AI technologies spark debates on whether they will replace or augment human creativity and labor. The victory of the AI-created Théâtre d'Opéra Spatial at the 2022 Colorado State Fair highlighted this issue, mixing excitement and concern over advanced Generative AI models. Research into the artist-AI relationship, such as Mazzone and Elgammal's study \cite{mazzone2019art}, argues that artists play a crucial role in the creative process with Generative AI, advocating for evaluating AI art by considering the artistic journey as well as the outcome. 
Oppenlaender's paper \cite{oppenlaender2022creativity} adopted Rhodes' four P model of creativity (people, process, product, and press) to analyze artwork created with text-to-image generative models, suggesting that generative AI art should be evaluated not solely based on the physical output but also the holistic factors involved in the art-making process.  

This discussion aligns with conceptual art, emphasizing the artist's concept over traditional techniques and materials. This study posits that in Generative AI Art, the artist's ideas drive the creation, with machines realizing these concepts. It centers on the research question: How to understand creativity in the era of Generative AI in relation to Conceptual Art? To address this, the project offers an analytical framework for interpreting creativity in Generative AI Art, linking it with Conceptual Art, and presents an art installation that merges both to explore the origins of creativity in the art-making process. 
Overall, this research project has made the following contributions:
\begin{itemize}[noitemsep,topsep=0pt]
\item \textbf{An Analytical Framework to Understand Creativity:} A close comparison of Conceptual Art and Generative AI Art is conducted using the 5P model of creativity to understand Creativity.
\item \textbf{An Interactive Installation Practice:}  Equivalence is an interactive installation practice that connects Generative AI Art with Conceptual Art as a manifestation of the 5P model.
\item \textbf{An Exploration of Language And Art:} Language in this project is represented in four levels: text recognition, language features, 3D visualization generated by the features, and 2D interpretation.
\end{itemize}

\section{Creating Artwork with Image
Generative AI}
Conceptual art emerged in the 1960s against traditional forms of art that focused on the physical object, such as painting or sculpture. Artists of Conceptual Art usually emphasize the idea or concept behind the artwork, often involving the use of instructions or rules to generate an artwork. Similarly, artwork created with generative AI relies on machines and algorithms to create art, with artists setting up rules, prompts, and parameters for the AI to follow. Table. \ref{Tab:comparing} compares the process of creating conceptual art with the process of creating art using Image Generative AI, focusing on the role of artists, especially the creative input, in the art creation process. The comparison is conducted based on the following aspects: People, Procedure, Product, and Press (from the 4P model\cite{4P} of creativity introduced by Rhodes in 1961), and Purpose to explain how Conceptual Art and Generative AI art are interconnected and how artists' creativity is performing an irreplaceable role even when creating artwork with Generative AI.

Through the analysis, it can be seen that, first, artists’ creativity inputs happen throughout the creation process and are the actual drive of the Generation process. Second, artists can control the creation process from different perspectives, including text prompts and image prompts construction, models selection and fine-tuning, post-curation, etc. Similar to Conceptual Art, the evaluation of Generative AI artwork should be extended to the whole creative process instead of the physical output.

\vspace{-5mm}
\setlength{\belowcaptionskip}{-10pt} 
\begin{table}[ht]
  \caption{Comparing Conceptual Art and Art Created with Generative AI}
  \label{Tab:comparing}
  \centering
  \begin{tabular}{p{1.4cm}p{6.5cm}p{6.5cm}}
   \hline
  \textbf{Perspective} &\textbf{Conceptual Art} &\textbf{Art Created with Generative AI}\\
   \hline
   Purpose & Conceptual art emphasizes the idea or concept over the physical creation, using instructions, rules, and collaboration to highlight art's intellectual value and thought process, rather than just craftsmanship. & Midjourney \cite{Midjourney} and Stable Diffusion \cite{StableDiffusion} aim to enhance imagination and democratize AI, allowing users to create images from text and manifest creativity regardless of technical skills. The generated output serves as a tangible representation and visualization of users’ creativity. \\

   People & 
Artists design the concept, set the instructions, and establish the intellectual groundwork, prioritizing creative thinking, critical analysis, and conceptual communication. Craftsmen, guided by these directions, execute the work through their own interpretation and understanding. & Artists dictate ideas and aesthetic criteria to AI, which generates outputs to meet these standards. The AI's creativity is shaped by human programming, using its "knowledge" to align each output with the artist's instructions for subsequent creation steps. \\ 
   Procedure & 
In conceptual art, the development of concepts is paramount, overshadowing techniques, craftsmanship, or materials, which are seen as mere mediums. For instance, Sol LeWitt's "Wall Drawings" were created through written instructions specifying geometric shapes, patterns, and colors, with the actual execution delegated to assistants or technicians.~\cite{lewitt1967paragraphs}&
Fig. \ref{fig:creativeprocess} shows the creative process for making artwork with AI tools, inspired by A. Elgammal's diagram \cite{mazzone2019art}. Artists use text prompts or base images to guide the image generation, which are processed by an AI model with artist-set parameters. Refining prompts and adjusting parameters through iterations are common, leading to the curation of a final collection from multiple outputs. \\ 
   Product & 
The final product of conceptual art—whether as text, instructions, installations, performances, music, or other chosen mediums—serves to manifest and convey artists' ideas, emphasizing the originality and impact of the concepts themselves. & 
The product encompasses the algorithm's visual or auditory outputs and the conceptual framework guiding the artwork's design. Evaluations within this dimension consider the novelty, originality, and impact of the ideas, as well as the effectiveness of their conveyance and the aesthetic qualities inherent in the artwork.\\ 
Press (Fig. \ref{fig:scatterplot} ) & 5 reviews articles on conceptual artwork are collected.The words ”idea” and ”concept” are mentioned in evaluating the value and creativity of the artwork. These terms can be seen as the ”product” of Conceptual Art, emphasizing the conceptual exploration and intellectual aspects of the artwork. & 5 reviews articles on Generative AI Artwork are collected. "AI," "model," and tools like "Dalle" are frequently mentioned, showing an emphasis on algorithms, possibly even more than on the artists. Terms like "train" and "generate" are more common than those emphasizing the artist's role, such as "create" and "think."  \\
   \hline   
  \end{tabular}
\end{table}
\vspace*{-\baselineskip}

\begin{figure}
  \centering
  \includegraphics[width=12cm]{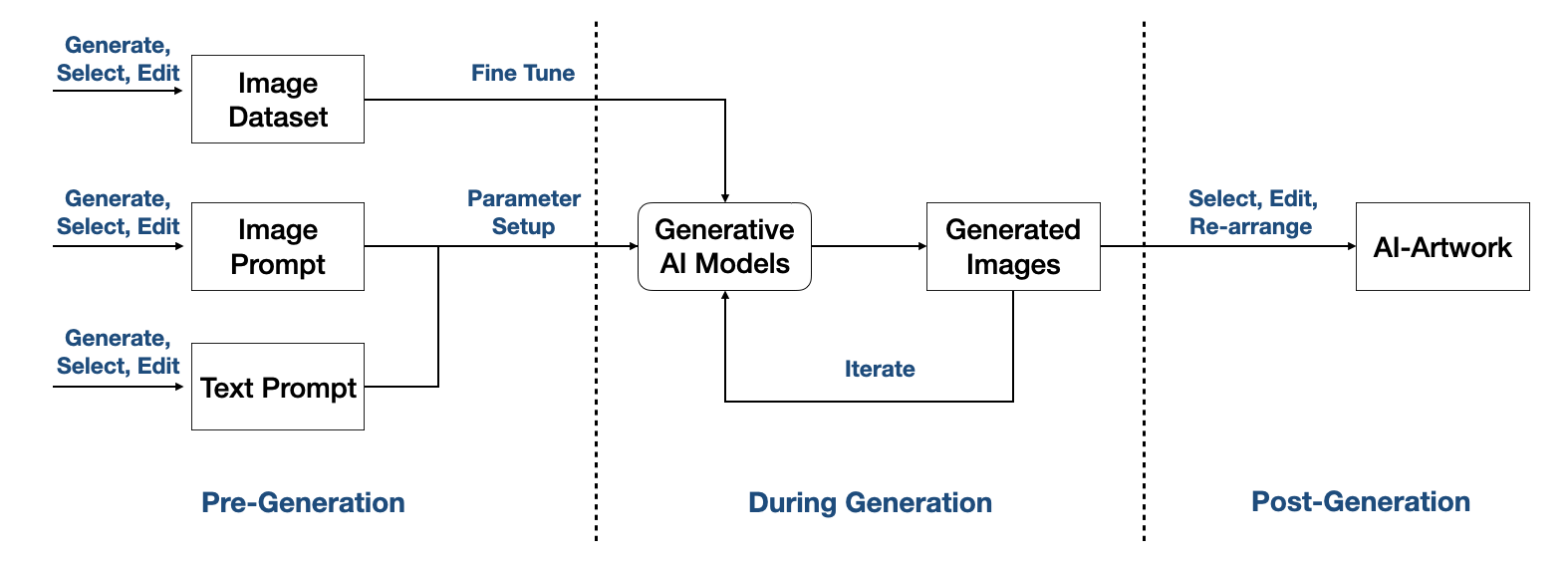} \centering
  \caption{Artists’ roles in creating images with Image Generative AI}
  \label{fig:creativeprocess}
\end{figure}

\vspace{5mm}
\section{Practice: Equivalence}

\begin{figure}
  \centering
  \includegraphics[width=10cm]{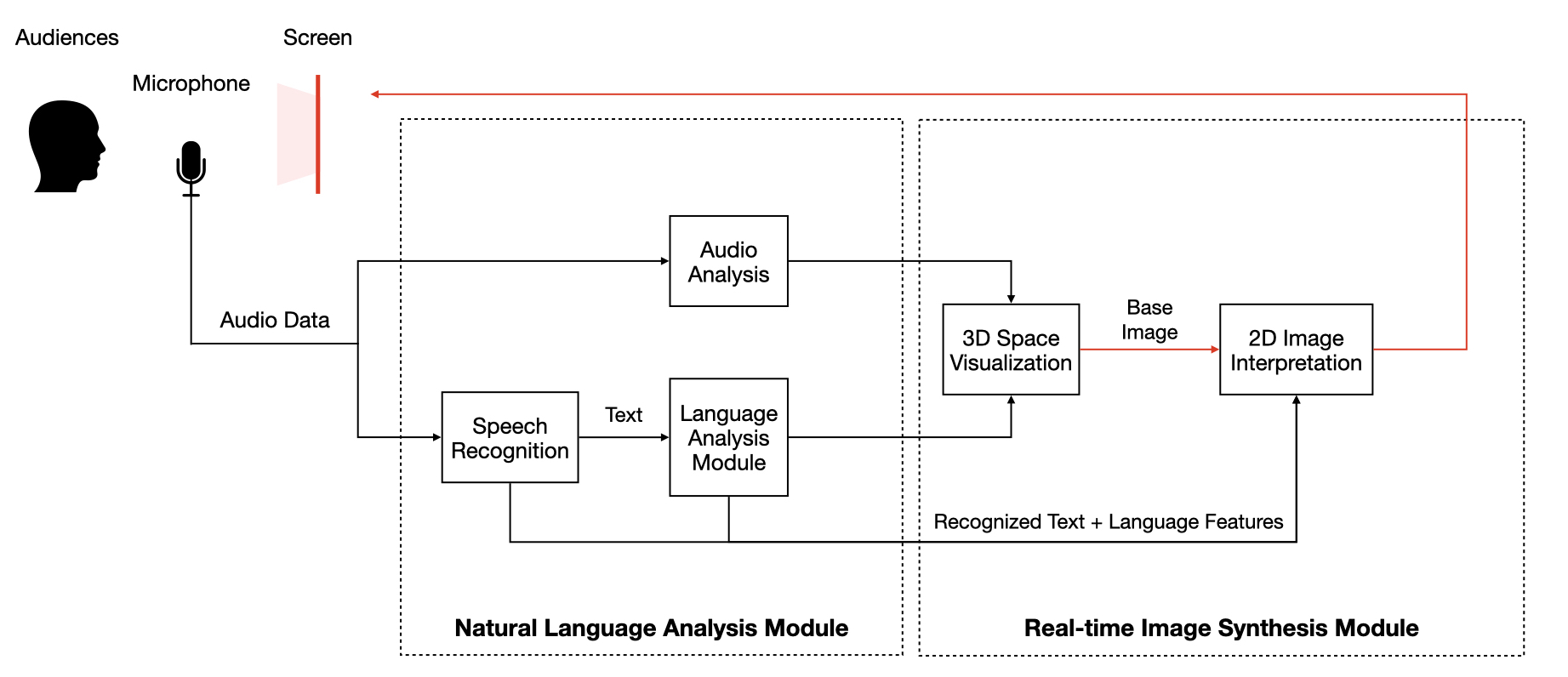} \centering
  \caption{A system design flowchart showing Equivalence’s workflow}
  \label{fig:flowchart}
\end{figure}

"Equivalence" is an interactive installation that visualizes spoken language in real-time using natural language processing and generative AI. Inspired by how utterances are structured, it converts speech into 3D spatial structures, presented in a dynamic Chinese scroll painting style.  It explores the relationship between grammar (structure) and words (content) in the construction of spoken languages, and maps this relationship to the artwork-creating process using language analysis to construct image composition and then the stable diffusion model to construct image textures. Equivalence is constructed with three important components: the natural language analysis part for recognizing and analyzing microphone input to construct the composition of the base image, the real-time image synthesis part for adding complexity and
texture to the base image, and the interaction part where audiences interact with the
installation and get real-time feedback. Therefore, Equivalence comes with the methodology of advanced machine intelligence system design, image synthesis, visualization techniques, and Human-Centered Interaction. Fig. \ref{fig:flowchart} shows the structure of the Equivalence system, including the microphone and the screen in the physical installation, the Natural Language Analysis Module for speech information capture, and the Real-Time Image Synthesis Module that consists of two image generation engines (one for base image generation from language analysis result and the other for final visualization generation using Stable Diffusion Model). 

\section{Discussion: Analyzing from the 5P model}
The purpose of the usage of Generative AI is to manifest the idea of combining semantic information and language features to construct visualization from speech and interpret language from different levels. The artwork generation process is entirely controlled and guided by the artists' design. Image prompts are generated using predefined parameters and mapping rules that we have established. Text prompts are constructed based on the text itself and guided by language features, following rules defined by the artists. Once generated, the images undergo further processing and rearrangement to form a long visualization. In this setup, artists, audiences, and machines are essential for the interactive installation. While machines are set up by programmers, artists fully control and direct them to realize the artistic concept and aesthetics. Audience participation, through voice input, also plays a crucial role. Together, these elements create a multi-screen interactive installation in an open exhibition space.
\section{Conclusion}
This research project conducts an investigation of artists' roles when collaborating with Generative AI. Through the analytical comparison and the creation of artwork, this research aims to demonstrate the effectiveness of the 5P model of creativity and provide strong support for the case that artists' ideas serve as the primary driving force behind Generative AI Artwork. It highlights creativity's role throughout the art-making process, where artists guide aesthetics and concepts, and AI materializes these ideas. The research advocates for evaluating such collaborative artworks by considering the entire creative process, not just the final output, underscoring the synergy between artists and technology.
\bibliographystyle{ACM-Reference-Format}
\bibliography{main}


\begin{thebibliography}{6}


\ifx \showCODEN    \undefined \def \showCODEN     #1{\unskip}     \fi
\ifx \showDOI      \undefined \def \showDOI       #1{#1}\fi
\ifx \showISBNx    \undefined \def \showISBNx     #1{\unskip}     \fi
\ifx \showISBNxiii \undefined \def \showISBNxiii  #1{\unskip}     \fi
\ifx \showISSN     \undefined \def \showISSN      #1{\unskip}     \fi
\ifx \showLCCN     \undefined \def \showLCCN      #1{\unskip}     \fi
\ifx \shownote     \undefined \def \shownote      #1{#1}          \fi
\ifx \showarticletitle \undefined \def \showarticletitle #1{#1}   \fi
\ifx \showURL      \undefined \def \showURL       {\relax}        \fi
\providecommand\bibfield[2]{#2}
\providecommand\bibinfo[2]{#2}
\providecommand\natexlab[1]{#1}
\providecommand\showeprint[2][]{arXiv:#2}

\bibitem[LeWitt(1967)]%
        {lewitt1967paragraphs}
\bibfield{author}{\bibinfo{person}{Sol LeWitt}.} \bibinfo{year}{1967}\natexlab{}.
\newblock \showarticletitle{Paragraphs on conceptual art}.
\newblock \bibinfo{journal}{\emph{Artforum}} \bibinfo{volume}{5}, \bibinfo{number}{10} (\bibinfo{year}{1967}), \bibinfo{pages}{79--83}.
\newblock


\bibitem[Mazzone and Elgammal(2019)]%
        {mazzone2019art}
\bibfield{author}{\bibinfo{person}{Marian Mazzone} {and} \bibinfo{person}{Ahmed Elgammal}.} \bibinfo{year}{2019}\natexlab{}.
\newblock \showarticletitle{Art, creativity, and the potential of artificial intelligence}. In \bibinfo{booktitle}{\emph{Arts}}, Vol.~\bibinfo{volume}{8}. MDPI, \bibinfo{pages}{26}.
\newblock


\bibitem[Midjourney(2022)]%
        {Midjourney}
\bibfield{author}{\bibinfo{person}{Midjourney}.} \bibinfo{year}{2022}\natexlab{}.
\newblock \bibinfo{title}{Introducting Midjourney}.
\newblock
\newblock
\urldef\tempurl%
\url{https://www.midjourney.com/home/}
\showURL{%
\tempurl}


\bibitem[Oppenlaender(2022)]%
        {oppenlaender2022creativity}
\bibfield{author}{\bibinfo{person}{Jonas Oppenlaender}.} \bibinfo{year}{2022}\natexlab{}.
\newblock \showarticletitle{The Creativity of Text-to-Image Generation}. In \bibinfo{booktitle}{\emph{Proceedings of the 25th International Academic Mindtrek Conference}}. \bibinfo{pages}{192--202}.
\newblock


\bibitem[Rhodes(1961)]%
        {4P}
\bibfield{author}{\bibinfo{person}{Mel Rhodes}.} \bibinfo{year}{1961}\natexlab{}.
\newblock \showarticletitle{An analysis of creativity}.
\newblock \bibinfo{journal}{\emph{The Phi delta kappan}} \bibinfo{volume}{42}, \bibinfo{number}{7} (\bibinfo{year}{1961}), \bibinfo{pages}{305--310}.
\newblock


\bibitem[Rombach et~al\mbox{.}(2022)]%
        {StableDiffusion}
\bibfield{author}{\bibinfo{person}{Robin Rombach}, \bibinfo{person}{Andreas Blattmann}, \bibinfo{person}{Dominik Lorenz}, \bibinfo{person}{Patrick Esser}, {and} \bibinfo{person}{Björn Ommer}.} \bibinfo{year}{2022}\natexlab{}.
\newblock \bibinfo{title}{High-Resolution Image Synthesis with Latent Diffusion Models}.
\newblock
\newblock
\showeprint[arxiv]{2112.10752}~[cs.CV]


\end{thebibliography}

\appendix

\section{Additional Images}
\begin{figure} 

  \centering
  \includegraphics[width=12cm]{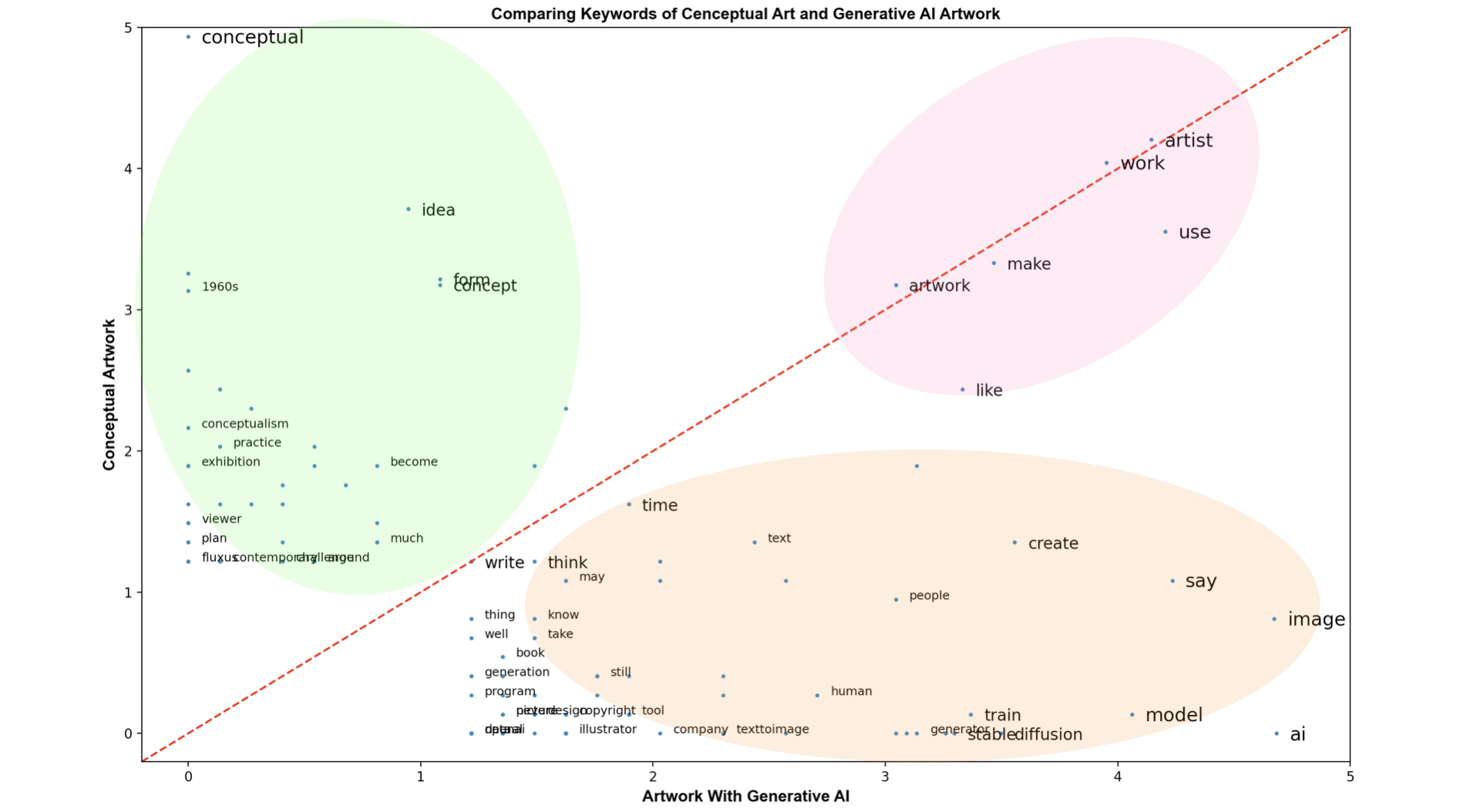} \centering
  \caption{Scatterplot comparing words frequency in reviews of Conceptual Art and Generative AI Art}
  \label{fig:scatterplot}

\end{figure}

\begin{figure}

  \centering
  \includegraphics[width=11cm]{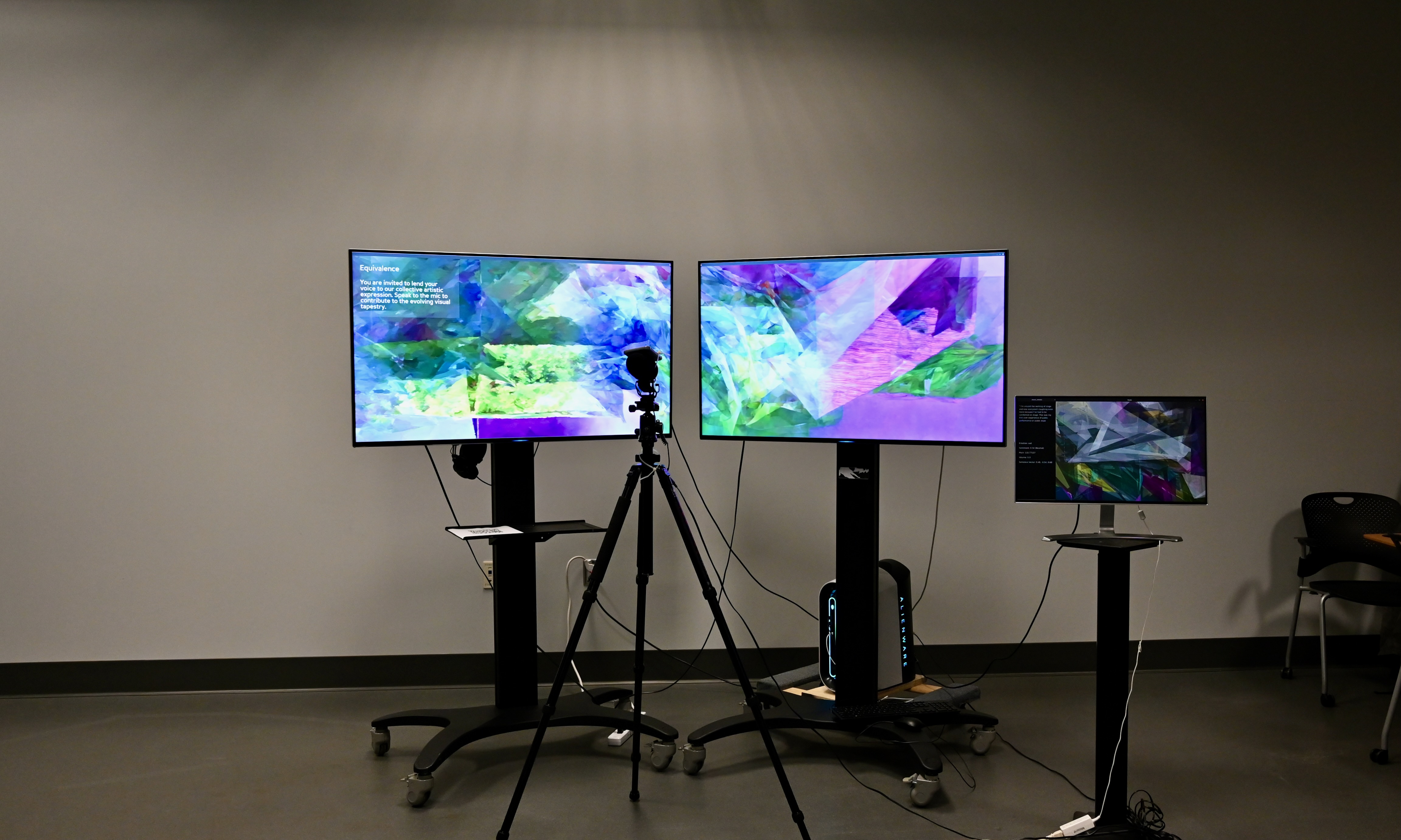} \centering
  \caption{ Installation on MAT End of Year Show with a simplified screens setup.}
  \label{fig:settp}

\end{figure}

Fig. \ref{fig:scatterplot} and Fig. \ref{fig:settp}are two additional images in support of explaining the work. 

Fig. \ref{fig:scatterplot} visualize the frequency of words appearing in the collected articles of the reviews of conceptual art and artwork created with Generative AI, of which the x-axis represents the frequency of words in the conceptual art articles and the y-axis represents the frequency of words in the Generative AI articles. Words that appeared more than 8 times across the 10 articles are included in the visualization. The red line connects the diagonal, and the proximity of a word’s point to the line indicates the similarity in frequency between the two categories. Words
appearing above or to the left of the red line suggest a higher frequency in articles about
Generative AI, while words appearing below or to the right of the line indicate a higher
frequency in Conceptual Art articles. 

Equivalence was exhibited as part of MAT End of Year Show. Fig. \ref{fig:settp} was the setup from the show. During the show, audiences were
invited to activate the visualization with their voices and experience how languages could
be interpreted and represented differently.

\end{document}